\documentclass[12pt]{article}
\usepackage[T1]{fontenc}
\usepackage[latin9]{inputenc}
\usepackage[letterpaper]{geometry}
\usepackage{units}
\usepackage{amsmath}
\usepackage{fixltx2e}
\usepackage{graphicx}
\usepackage{setspace}
\makeatletter

\providecommand{\tabularnewline}{\\}

\newcommand{\lyxaddress}[1]{
\par {\raggedright #1
\vspace{1.4em}
\noindent\par}
}

\@ifundefined{date}{}{\date{}}
\usepackage{overcite}

\makeatother

\begin{document}
\begin{singlespace}

\title{\noindent {\Large{Performance of transducers with segmented piezoelectric
stacks using materials with high electromechanical coupling coefficient}}}
\end{singlespace}

\author{Stephen C. Thompson%
\thanks{electronic address: steve.thompson@psu.edu%
}, Richard J. Meyer and Douglas C. Markley}

\maketitle

\lyxaddress{Applied Research Laboratory MS2430, The Pennsylvania State University,
State College, Pennsylvania 16804-0030}

\vspace{1in}

Suggested running title: high $k_{33}$ segmented piezoelectric stacks

\date{submitted: January 30, 2013}

\pagebreak{}
\begin{abstract}
\noindent {\normalsize{Underwater acoustic transducers often include
a stack of thickness polarized piezoelectric material pieces of alternating
polarity interspersed with electrodes, bonded together and electrically
connected in parallel. The stack is normally much shorter than a quarter
wavelength at the fundamental resonance frequency, so that the mechanical
behavior of the transducer is not affected by the segmentation. When
the transducer bandwidth is less than a half octave, as has conventionally
been the case, stack segmentation has no significant effect on the
mechanical behavior of the device. However, when a high coupling coefficient
material such as PMN-PT is used to achieve a wider bandwidth, the
difference between a segmented stack and a similar piezoelectric section
with electrodes only at the two ends can be significant. This paper
investigates the effects of stack segmentation on the performance
of wideband underwater acoustic transducers, particularly tonpilz
transducer elements. Included is discussion of transducer designs
using single crystal piezoelectric material with high coupling coefficient
compared with more traditional PZT ceramics.}}{\normalsize \par}
\end{abstract}
PACS numbers: 43.38Fx, 43.378Ar

\noindent Keywords: piezoelectric

\pagebreak{}

\section{Introduction}

\begin{figure}[h]
\begin{centering}
\includegraphics[width=3in]{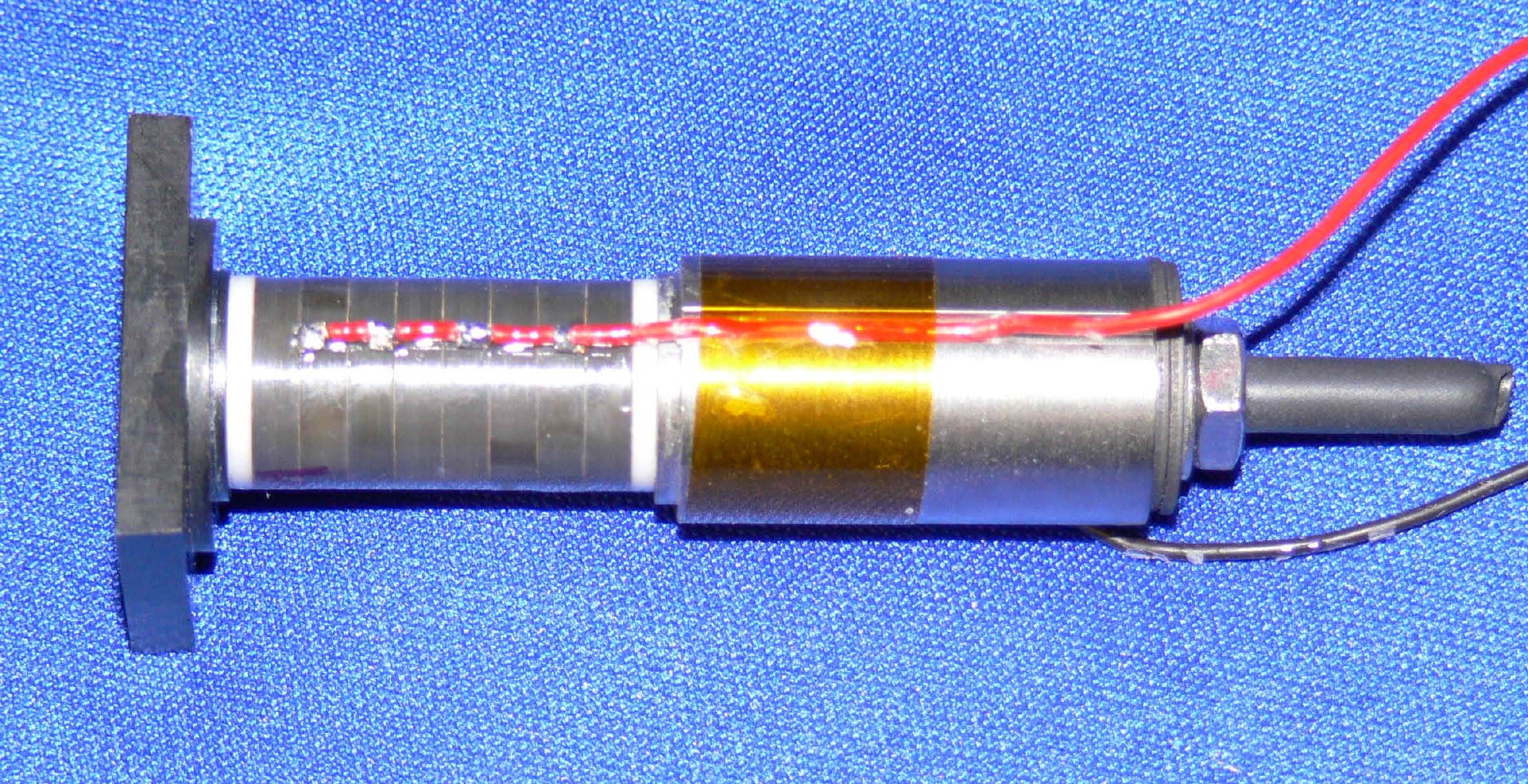}
\par\end{centering}

\caption{The tonpilz transducer element used in this study has a stack of eight
PMN-PT crystal rings. (color online)\label{fig:The-tonpilz-transducer}}
\end{figure}

Transducers based on piezoelectric ring stacks were conceived by Miller~\cite{miller-ring-stack}
and have been used at least since 1959. Since then, piezoelectric
sonar transducers are often constructed using a stack of thickness-polarized
piezoelectric rings or plates. An example of such a transducer is
shown in Figure~\ref{fig:The-tonpilz-transducer}. The segmentation
of the piezoelectric material into rings or plates may be done to
facilitate the manufacturing and polarization of the piezoelectric
pieces and/or to reduce the voltage needed to achieve the electric
field required for full output power. In either case, the stack is
assembled by alternating the polarization directions of adjacent rings
in the stack, and alternating the voltages on the electrodes between
the rings. The rings are stacked mechanically in series, but wired
electrically in parallel. In this way, at least for quasistatic operation,
all of the piezoelectric pieces are driven to expand and contract
simultaneously. 

\noindent 

\begin{figure}
\noindent \begin{centering}
\includegraphics[width=3in]{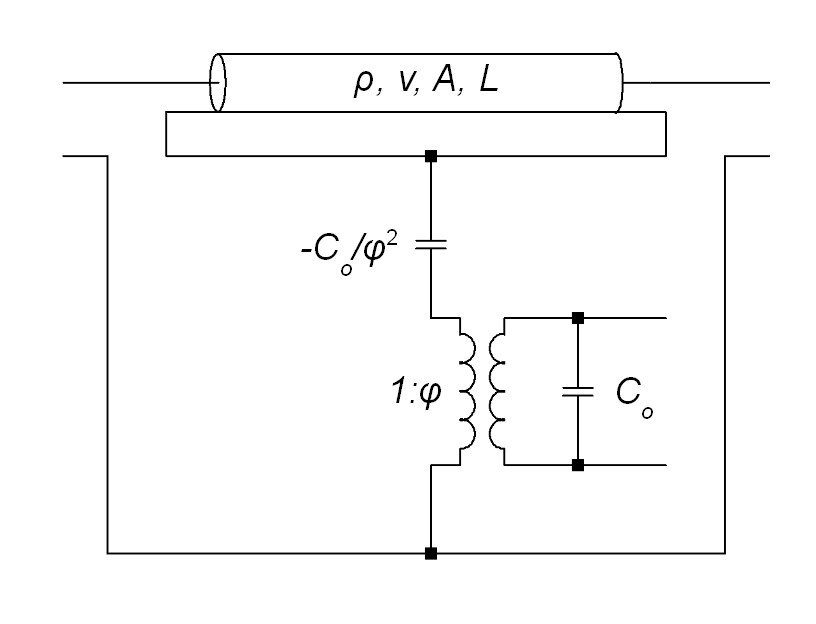}
\par\end{centering}

\caption{Analog circuit model for a single piezoelectric piece, after Redwood.
The mechanical domain includes the mechanical transmission line with
density $\rho$, sound speed $v$, cross sectional area $A$ and length
$L$. \label{fig:Redwood-ckt}}
\end{figure}

The first description of a ring stack assembly is by Miller\cite{miller-ring-stack}.
Previously, Mason had introduced a method of analysis and an analog
circuit for individual piezoelectric plates and tubes, initially using
a lumped parameter analog circuit~\cite{mason-IRE+BSTJ} and later
allowing longitudinal plane wave motion in the piezoelectric piece~\cite{mason}.
Redwood~\cite{redwood} recognized that the mechanical domain in
the Mason analog circuit could be represented as a mechanical transmission
line. Figure~\ref{fig:Redwood-ckt} shows the Redwood implementation
of the Mason analog circuit for a single piezoelectric piece. Martin
was the first to analytically describe the operation of the segmented
stack. He first provided an analysis of the performance of longitudinally
polarized piezoelectric tubes that includes effects of axial stress
in the tube walls~\cite{martin-tubes}. He then described the analysis
of the ring stack as an assembly of short tubes or rings with their
mechanical ports connected in cascade and their electrical ports connected
in parallel~\cite{martin-segmented,martin-segmented-33mode}.

\noindent 

The conclusion of the Martin analysis is that a piezoelectric stack
of $p$ identical rings, each of length $L$ 
\begin{enumerate}
\item has the mechanical behavior identical to that of a single long tube
of length $pL$ with slightly modified material parameters, and 
\item has the electrical behavior consistent with the $p$ short rings wired
electrically in parallel.
\end{enumerate}
Martin stated that his analysis assumes only that the rings have identical
dimensions and material properties, and that the length of each ring
is much less than a half wavelength in the piezoelectric material,
$L\ll\tfrac{\lambda}{2}$. These conclusions significantly simplify
the design process for a tonpilz transducer. The projector designer
can select the total stack length $pL$ to provide sufficient dynamic
displacement, and can separately determine the total number of rings
in the stack to assure the producibility of the rings, and to obtain
an appropriate electrical impedance that keeps the drive voltage level
required for full power operation to a moderate level. Wilson~\cite{Wilson}
and Stansfield~\cite{Stansfield} each give an example of the use
of the Martin conclusions. The unstated assumption of both authors
is that the segmentation of the piezoelectric stack does not affect
the mechanical behavior of the transducer.

Despite the acceptance and utility of the Martin conclusions, the
Appendix shows that there is both an error and an omission in the
Martin derivation that have been previously unrecognized, probably
because they introduce negligible error in the calculation when used
with materials that have an electromechanical coupling coefficient
less than about 0.75. That includes all of the piezoelectric materials
that were available to Martin, Wilson, Stansfield and others at the
time of their work. In the next section, however, those errors will
be shown to be important when using single crystal piezoelectric materials
that have electromechanical coupling coefficient higher than 0.75,
particularly when those materials are used to implement transducers
with higher power and wider bandwidth than is possible with PZT and
other conventional piezoelectric materials.

The error in Martin is his claim that his approximation holds whenever
the length of the individual ring in the stack is small, $kL\ll\pi$.
The appendix shows that it is the total length of the stack, rather
than the length of the individual ring, that must be small. For a
stack of $p$ pieces, the requirement is that $pkL\ll\pi$. By itself,
that is not a serious limitation, as most designers would expect the
total stack length to be significantly less than a wavelength in all
cases, in order to maintain a sufficient frequency separation from
the first ``higher mode'' mechanical resonance within the stack.
However, that situation needs to be reexamined in light of the significant
bandwidth that may be available with high coupling coefficient piezoelectric
materials.

The omission in Martin is the recognition from Equation \ref{eq:delta-n-2-approx}
in the Appendix that the error in the approximation is proportional
to $\nicefrac{n^{3}k_{33}^{2}}{\left(1-k_{33}^{2}\right)}$. This
means that the length constraint is relatively more important when
either the coupling coefficient or the number of stack segments is
high.

Of course, all of the statements above are qualitative and cautionary,
but provide no information on the limits of the approximation in any
specific case. The balance of this paper shows the effect on the analysis
of the particular transducer shown in Figure \ref{fig:The-tonpilz-transducer}.

The effects of stack segmentation on transducer performance will be
shown below, both analytically and with data from the transducer shown
in Figure \ref{fig:The-tonpilz-transducer}. The effects are apparent
in an analog transmission line model within the limitations of accuracy
of such a model. 

Note that the information in this paper should not be interpreted
as a negative report on materials with high electromechanical coupling.
Materials with high coupling enable transducer designs that are difficult
or impossible to achieve with other piezoelectric materials\cite{Sherlock&Meyer,Shear-mode}.
The relevant point is that the design methods and design intuitions
that have served well in the past may need to be revised to take full
advantage of properties of the new materials.

\section{Analysis}

The initial calculations of the element performance will be shown
with an analog circuit model using plane wave transmission lines to
model the mechanical properties of the piezoelectric stack. While
the longitudinal models are \emph{not }sufficient to fully describe
the behavior of the transducer hardware, they do clearly show the
effects of segmentation and are a simpler framework for understanding
the effects. A full FEA of the transducer is then used to include
the full three dimensional effects and for comparison with the experimental
hardware.

\subsection{Analysis Using a Plane Wave Model}

The initial calculations of the element performance will be shown
using an analog plane wave model implemented in the SPICE~\cite{SPICE,Nagel}
computer code using methods that were first described by Leach~\cite{Leach,Leach2}.
Each material piece in the transducer assembly is modeled as a mechanical
transmission line with constant area. Electromechanical coupling in
the piezoelectric material is modeled with controlled sources in SPICE
in the manner described by Leach~\cite{Leach2}. The interconnections
of material pieces follow the conventions of analog circuits using
the impedance analogy. The present work has been performed using the
LTspice code\cite{LTSpice}, although several other SPICE versions
are known to be suitable\cite{Other-SPICE}.

\begin{figure*}
\begin{centering}
\includegraphics[width=5.5in]{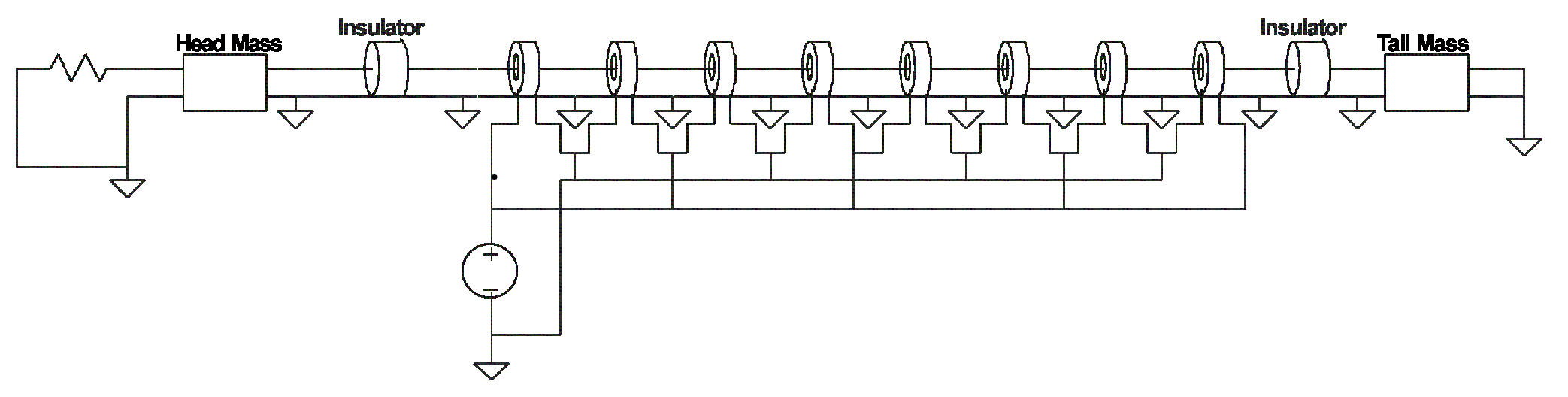}
\par\end{centering}

\caption{The plane wave model of the transducer is an interconnection of mechanical
transmission lines. The electrodes of the piezoelectric pieces are
connected electrically in parallel. \label{fig:Interconnection-8-pieces} }
\end{figure*}

The simplified transducer model is shown in Figure \ref{fig:Interconnection-8-pieces}.
This shows the interconnection of the mechanical pieces and the electrical
terminals of the transducer in a schematic-like graphic. The eight
piezoelectric rings at the center of the diagram are modeled individually
using the Leach~\cite{Leach2} implementation of the Redwood~\cite{redwood}
analog circuit to avoid any possibility of errors in the Martin approximation
described in the Appendix being present in this analysis. A thin insulator
is placed at each end of the piezoelectric stack. The head and tail
masses are connected to the insulators at opposite ends of the stack.
The head mass is terminated with a radiation resistance whose value
is $\rho cA$, where $\rho$ is the density of the medium, $c$ is
the sound speed in the medium, and $A$ is the radiating area. This
is a reasonable approximation if the element is used in a large array.
The back end of the tail has a zero force boundary condition, analogous
to an electrical connection to zero potential. The full analog circuit
implementation also includes a large number of ground terminations
in the electrical and mechanical domains to meet the SPICE requirement
that every circuit loop have a DC reference to ground. 

\begin{figure}
\begin{centering}
\includegraphics[width=3in]{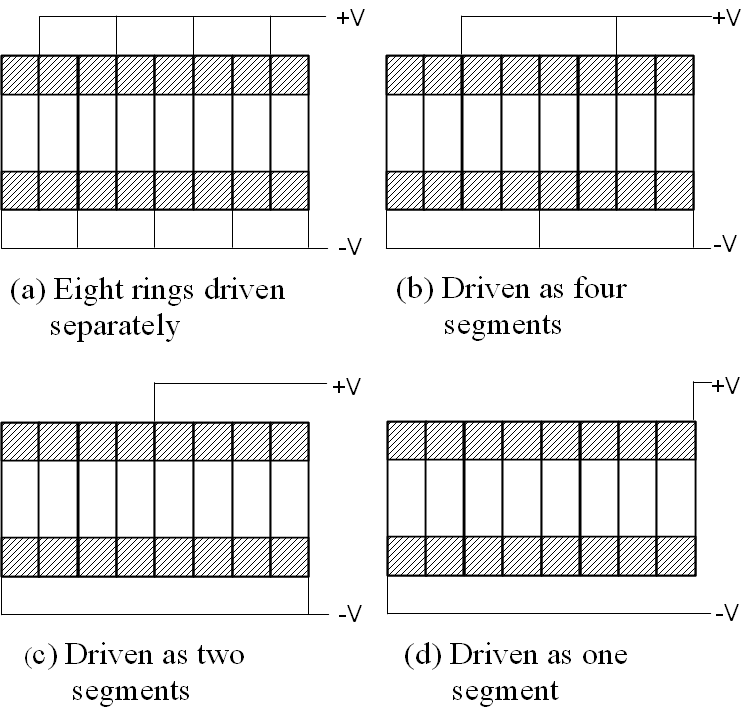}
\par\end{centering}

\caption{The piezoelectric stack is driven in four configurations, as eight
segments, four segments, two segments and one segment. Electrodes
remain in place for the eight rings, but not all are connected, as
shown.\label{fig:The-piezoelectric-stack}}
\end{figure}

For comparison, three other transducer designs were also analyzed,
with the piezoelectric section segmented into four pieces, two pieces,
and one piece as shown in \ref{fig:The-piezoelectric-stack}. In each
case, the individual piece length in the model was adjusted so that
the total length of the piezoelectric stack is the same as the length
of the eight piece stack in the model of Figure \ref{fig:Interconnection-8-pieces}.
The electrical connection of the piezoelectric rings is in parallel.
The rings are assembled with alternating polarization directions,
so that they are all driven to expand and contract in phase. Glue
joints and electrodes are not included in the model. The small added
compliance of the glue joints may explain a part of the difference
in resonance and antiresonance frequencies between measurement and
model that are seen in Section \ref{sub:Analysis-with-FEA}. A stress
rod that is present in the transducer element is not included in the
model at this stage of the analysis. In the actual transducer, the
stress rod adds very little stiffness to the assembly because it is
attached with a highly compliant spring washer at the tail. 

\noindent 

\noindent 

The analysis described above was performed using piezoelectric material
parameters for PZT-4 ceramic, and again using material parameters
appropriate for the high coupling PMN-PT single crystal material.
Values for the material parameters for this analysis were taken from
Sherman and Butler~\cite{Sherman&Butler}. The dimensions of the
piezoelectric pieces with the two materials were the same, and no
attempt was made to compensate for the difference in resonance frequency
caused by the different mechanical properties of the two materials. 

\noindent 
\begin{figure}
\noindent \begin{centering}
\includegraphics[width=3.375in]{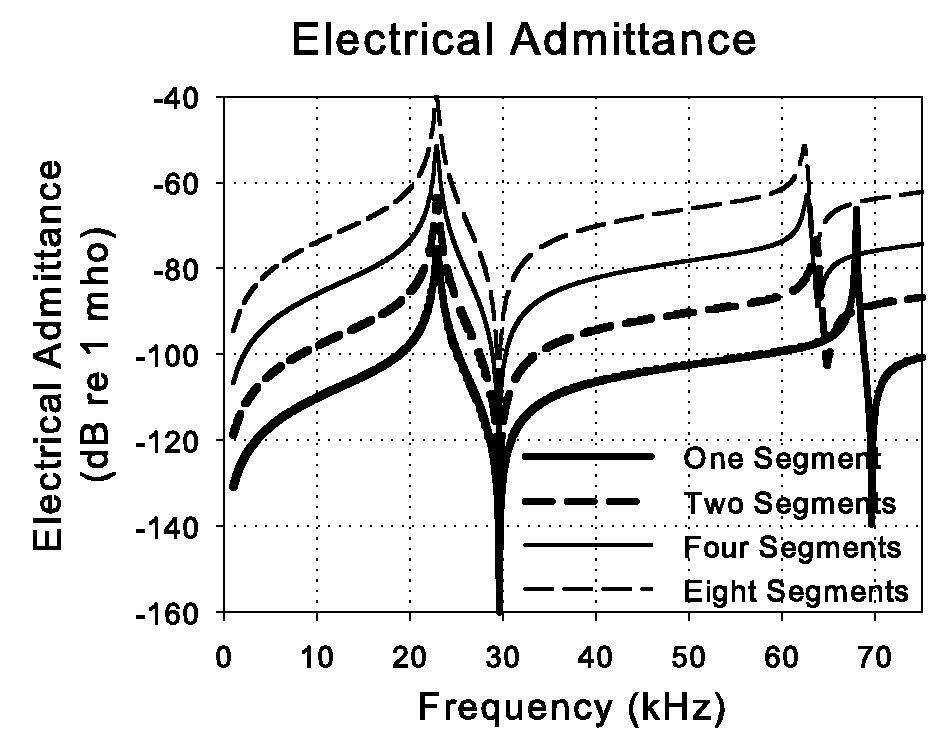}
\par\end{centering}

\caption{The electrical admittance for transducers built with PZT-4 piezoelectric
material do not vary significantly in the frequency region of the
primary resonance and antiresonance .\label{fig:The-electrical-admittance-PZT}}
\end{figure}

The electrical admittance curves calculated from the analog model
with air loading for the four PZT-4 ceramic stacks are shown in Figure
\ref{fig:The-electrical-admittance-PZT}. The 12 dB level difference
between curves at low frequencies is caused by the different segmentation
levels of the stack. There is a factor of four difference in stack
impedance when the number of rings is doubled and the ring thickness
is halved. The resonance frequencies and the antiresonance frequencies
calculated for the four transducers each differ by less than 0.5\%.
Note that at the second resonance, there is a noticeable difference
in resonance frequency between the unsegmented case and all of the
segmented cases. However, this difference is small enough that it
would likely not be recognized unless the segmented and unsegmented
curves were compared in this way, and even if it were noticed, it
would not be considered important. In PZT transducers, the second
resonance is always far enough above the primary resonance that it
has no noticeable impact on the performance in the primary band.

\noindent 
\begin{figure}
\noindent \begin{centering}
\includegraphics[width=3.375in]{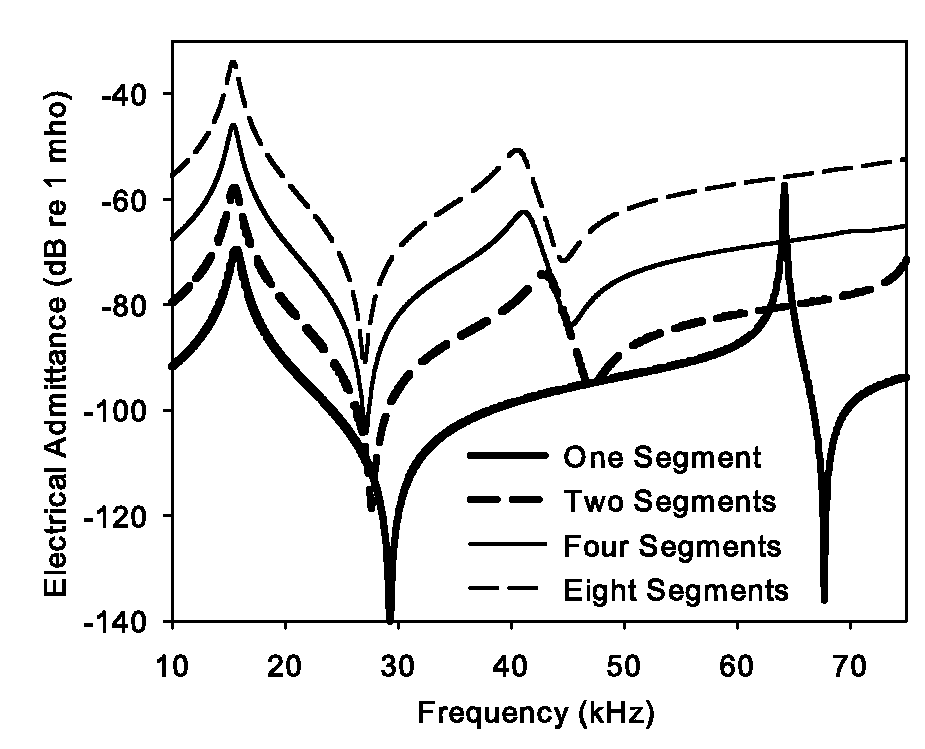}
\par\end{centering}

\caption{Electrical admittance of the four segmented stacks using PMN-PT show
greater differences than with PZT in the frequency region of the primary
antiresonance, and especially at the higher mode resonance.\label{fig:Electrical-admittance-pmnpt}}
\end{figure}

The electrical admittance curves with air loading for the four PMN-PT
single crystal stacks are shown in Figure \ref{fig:Electrical-admittance-pmnpt}.
The 12 dB level difference between curves at low frequencies is again
present due to the segmentation. The calculated resonance frequencies
of the four transducers do not vary significantly. However, the antiresonance
frequencies differ noticeably, and the second mode resonance frequencies
vary dramatically. Again, the difference is between the unsegmented
piece and all of the segmented stacks. The change in antiresonance
frequency reduces the effective coupling coefficient for the transducer
with segmentation by a bit over 2\%. The numerical values for these
frequencies and the effective electromechanical coupling coefficient
calculated from them are given in Table 1. The reduction in the frequency
of the second mode resonance with segmentation is quite noticeable,
now being less than a third octave above the antiresonance frequency.
This difference may be considered important in some applications,
as it may affect the operating bandwidth of the device. This will
be seen by looking at the Transmitting Voltage Response (TVR).

\noindent 
\begin{figure}
\noindent \begin{centering}
\includegraphics[width=3.375in]{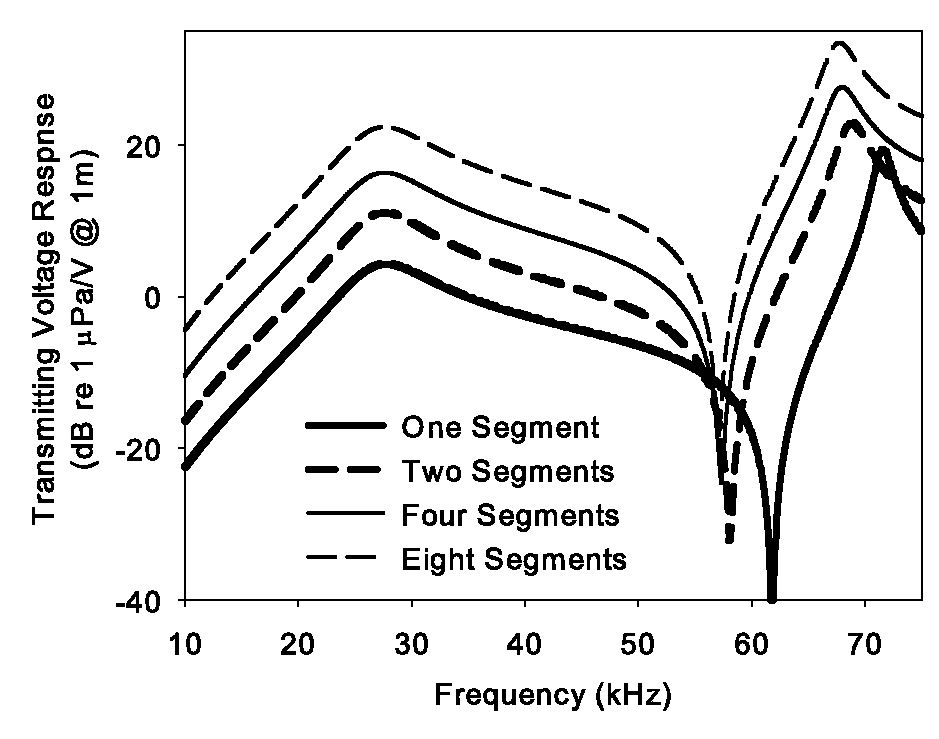}
\par\end{centering}

\noindent \caption{TVRs for the four transducer implementations using PZT-4.\label{fig:TVR-PZT}}
\end{figure}

The far field SPL for the element can be calculated approximately
as 
\begin{equation}
SPL=\dfrac{\omega\rho A}{2\pi R}\mid v_{h}\mid\label{eq:far-field-SPL}
\end{equation}
where $A$ is the area of the head, and $R$ is the measurement distance
and $v_{h}$ is the radiating velocity of the head. Thus the approximate
TVR for the element can be calculated from Equation \ref{eq:far-field-SPL}
with $R=1\mathrm{m}$, where the head velocity is calculated from
the analog model using a 1~V drive signal. The TVR calculated in
this manner with the radiation load of $\rho cA$ for the four implementations
of the transducer using PZT-4 are shown in Figure \ref{fig:TVR-PZT}.
The difference in the level of the TVR is due to the different electric
field in the piezoelectric material. For this calculation, a unit
voltage is applied to the element terminals, but the different amounts
of stack segmentation cause the voltage to be applied across piezoelectric
pieces of different thickness. The low frequency level difference
of 6 dB between adjacent curves is expected because the piezoelectric
piece thickness differs by a factor of two. 

\noindent 
\begin{figure}
\noindent \begin{centering}
\includegraphics[width=3.375in]{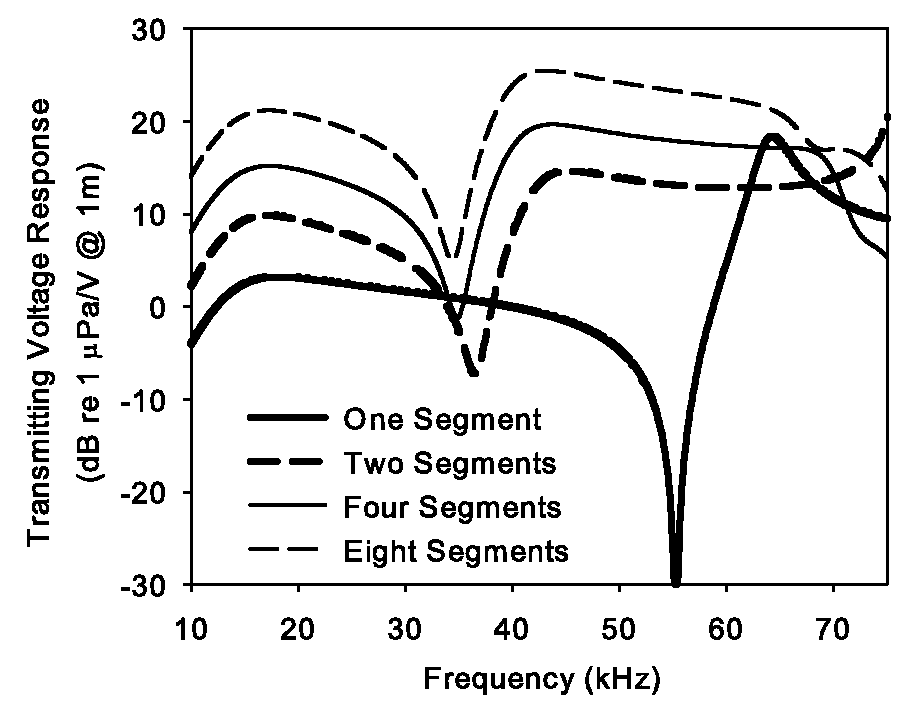}
\par\end{centering}

\caption{The TVR for transducers built with PMN-PT single crystal piezoelectric
material show that segmentation may cause significant variation in
operating bandwidth. \label{fig:The-TVR-PMNPT}}
\end{figure}

The TVR calculated in this manner with the radiation load of $\rho cA$
for the four implementations of the transducer using PMN-PT are shown
in Figure \ref{fig:The-TVR-PMNPT}. Again, the difference in the level
of the TVR at low frequencies is due to the different electric field
in the piezoelectric material. Using nearly any measure of operating
bandwidth, the bandwidth available with the high coupling material
is less when the stack is segmented than with the unsegmented stack. 

Note that the fractional bandwidth with high coupling material, even
with the segmentation, is still higher than the fractional bandwidth
using PZT. Using the same definition of bandwidth, the PZT transducer
may operate from 20 kHz to 40 kHz, while the PMN-PT transducer would
operate from 10 kHz to 30 kHz. There may be additional advantages
with the PMN-PT in that it operates with the same absolute bandwidth
at a lower center frequency, and that the lower frequency operation
is achieved with no increase in size. However, it remains true that
high coupling material without segmentation has an even greater bandwidth
with a relatively flat TVR.

\begin{table}
\begin{tabular}{|c|c|c|c|c|}
\hline 
\# Segments & 1 & 2 & 4 & 8\tabularnewline
\hline 
\hline 
\multicolumn{5}{|c|}{Using PMN-PT with $k_{33}=0.87$}\tabularnewline
\hline 
\hline 
$f_{r}$(Hz) & 15671 & 15520 & 15456 & 15434\tabularnewline
\hline 
$f_{a}$(Hz) & 29286 & 27650 & 27176 & 27044\tabularnewline
\hline 
$k_{eff}$(Hz) & 0.845 & 0.828 & 0.823 & 0.821\tabularnewline
\hline 
\multicolumn{5}{|c|}{Using PZT with $k_{33}=0.70$}\tabularnewline
\hline 
\hline 
$f_{r}$(Hz) & 27464 & 27388 & 27368 & 27368\tabularnewline
\hline 
$f_{a}$(Hz) & 35460 & 35239 & 35166 & 35142\tabularnewline
\hline 
$k_{eff}$(Hz) & 0.633 & 0.629 & 0.627 & 0.627\tabularnewline
\hline 
\end{tabular}

\caption{Measurable properties calculated for transducer elements made with
piezoelectric material with several levels of segmentation.\label{tab:Table}}
\end{table}

\subsection{Analysis with FEA\label{sub:Analysis-with-FEA}}

The previous predictions were based on a one dimensional model and
show the character of the differences between moderate and high coupling
materials with stack segmentation. However it does not model the fully
realistic motion within the transducer that includes axial motion
at least in the piezoelectric stack and the possibility of non-axially
symmetric motions such as would occur with a flexural resonance in
the head or other parts. The theory was extended to three dimensional
modeling using the ATILA finite element code. The geometry created
in the model matched the PMN-PT single crystal tonpilz element shown
in Figure \ref{fig:The-tonpilz-transducer}. The piezoelectric stack
was modeled in two ways, first with eight rings having alternating
polarity and wired as shown in Figure \ref{fig:The-piezoelectric-stack}(a),
and separately as a single piezoelectric piece wired as shown in Figure
\ref{fig:The-piezoelectric-stack}(d). Corresponding measurements
on this tonpilz element were also done in two ways. First, the motor
section was built using traditional methods with eight rings stacked
with alternating polarity separated with electrode shims as in Figure
\ref{fig:The-piezoelectric-stack}(a). This stack configuration is
labeled as \textquotedblleft{}parallel\textquotedblright{} in the
following figures as the rings are wired electrically in parallel.
The stack configuration was then altered in such a way as to connect
the rings in series. This was accomplished by reverse polarizing the
even rings in the stack thereby providing a continuous polarity direction
for the rings in the stack. The electrode tabs were then removed from
the intermediate electrodes, and the electrical potential was applied
across the stack as if it were one solid piece in Figure \ref{fig:The-piezoelectric-stack}(d).
This configuration is labeled \textquotedblleft{}series\textquotedblright{}
in the figures that follow and it is compared to a finite element
simulation that models this geometry.

This ``series'' configuration of the stack is similar to the physical
condition of a single piece, but not identical to it. A single piezoelectric
piece, or a stack of rings with no metal electrode in place, would
allow the electric potential to vary axially at the electrode positions.
With the metal electrodes in place, the potential is held constant
across this electrode surface. However the calculated results of the
FEA model matched the conditions with the conductors in place, so
the conditions during measurement are properly modeled. Furthermore,
the effect of the open electrodes in the stack is not expected to
be significant. The following section uses this model for the calculations
that are compared with measurements on a single transducer element
of Figure \ref{fig:The-tonpilz-transducer}.

\noindent 
\begin{figure}
\noindent \begin{centering}
\includegraphics[width=3.375in]{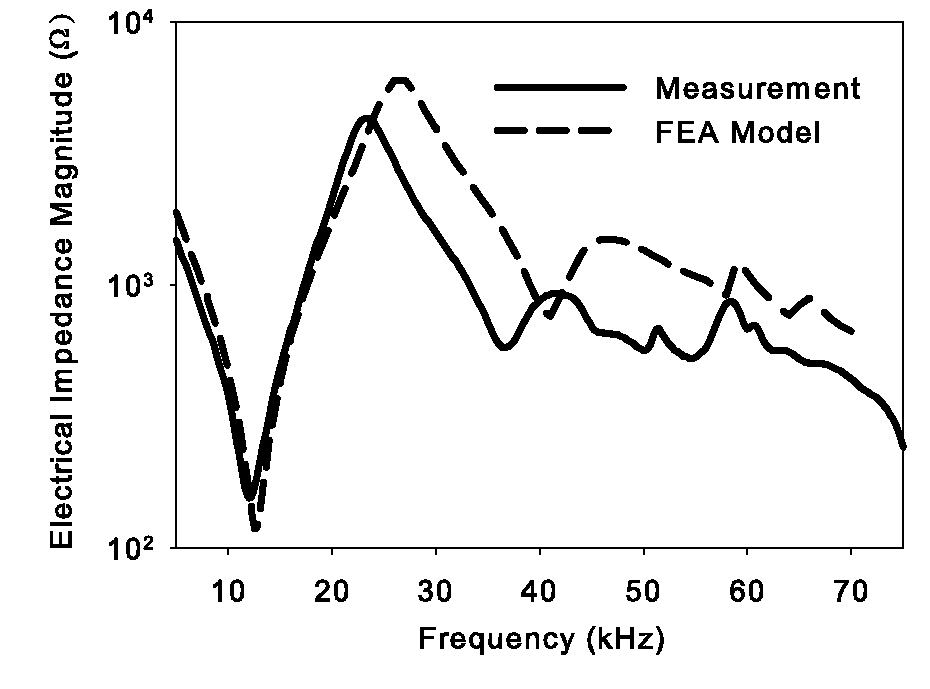}
\par\end{centering}

\noindent \begin{centering}
(a) parallel electrical connection
\par\end{centering}

\noindent \begin{centering}
\includegraphics[width=3.375in]{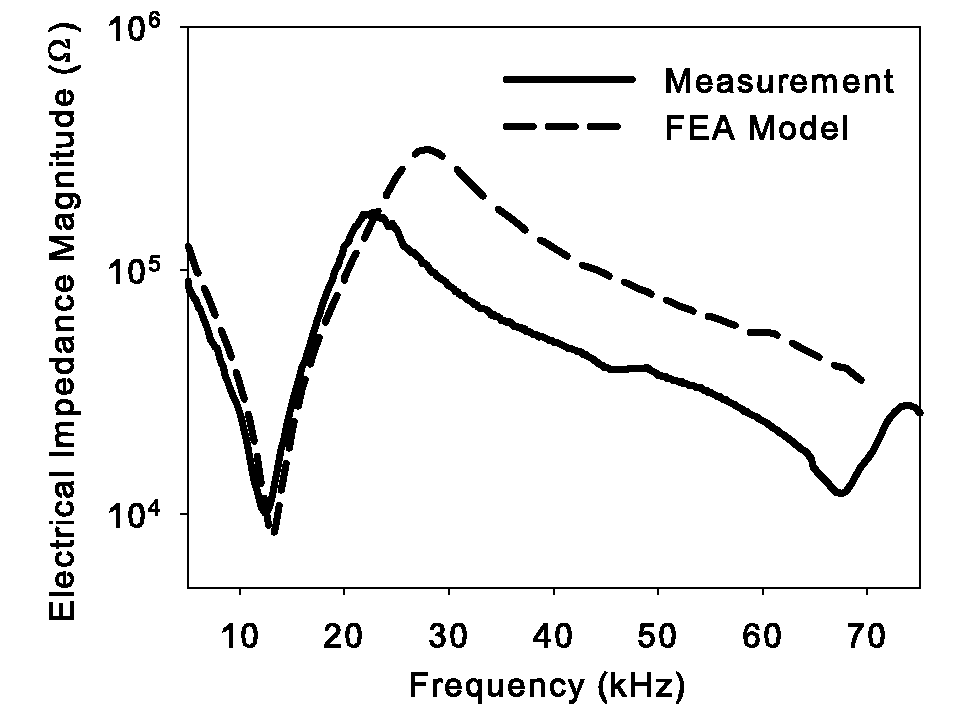}
\par\end{centering}

\noindent \begin{centering}
(b) series electrical connection
\par\end{centering}

\noindent \caption{In water electrical impedance magnitude for a single crystal PMN-PT
based tonpilz element that was both measured and modeled in a) parallel
and b) series stack arrangement.\label{fig:Zel-parallel-series}}
\end{figure}

\section{Verification with Measurement }

\begin{figure}
\noindent \begin{centering}
\includegraphics[width=3.375in]{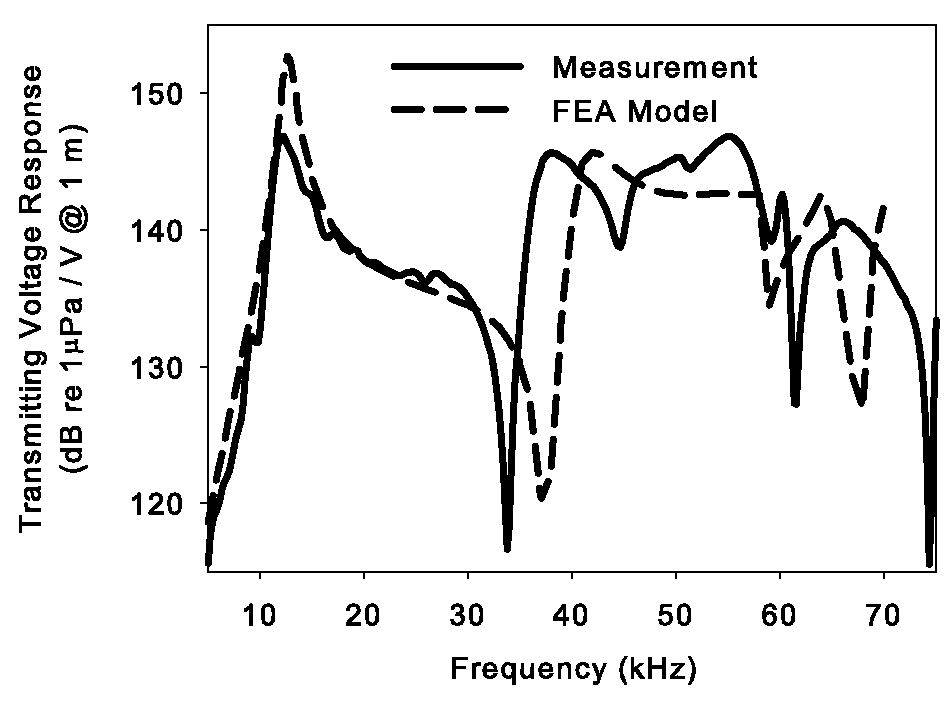}
\par\end{centering}

\noindent \begin{centering}
(a)
\par\end{centering}

\noindent \begin{centering}
\includegraphics[width=3.375in]{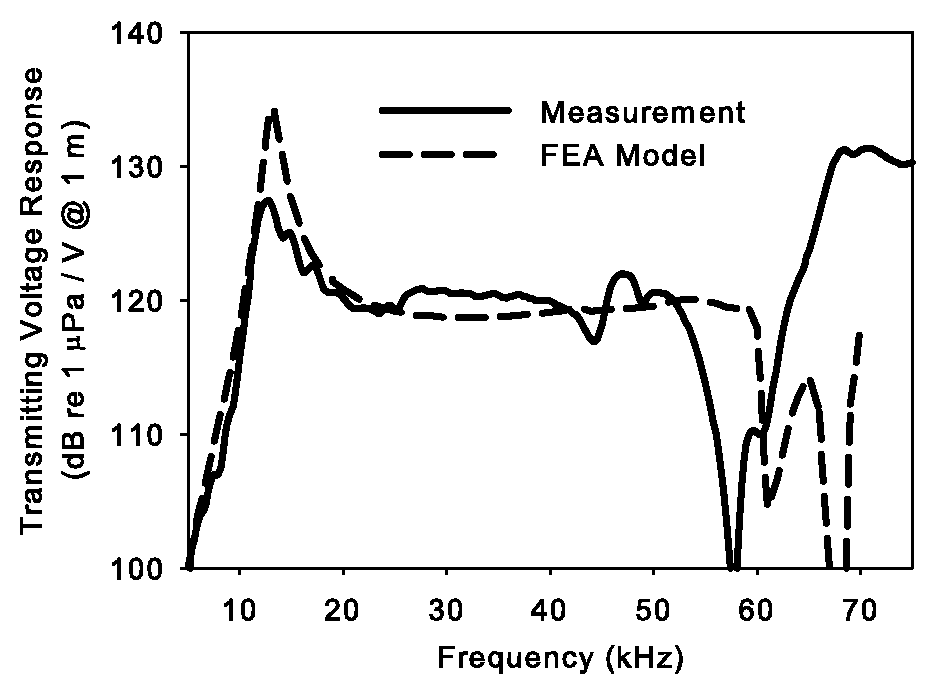}
\par\end{centering}

\noindent \begin{centering}
(b)
\par\end{centering}

\caption{The measured Transmitting Voltage Response in water for a single tonpilz
element of single crystal PMN-PT. The element was measured and modeled
in a) series and b) parallel stack arrangement.\label{fig:TVR-parallel-series}}
\end{figure}

Figures \ref{fig:Zel-parallel-series} and \ref{fig:TVR-parallel-series}
show the comparison of the model and measured results obtained for
the parallel and series stack arrangements. Figure \ref{fig:Zel-parallel-series}
is the electrical impedance magnitude for the element in water, and
Figure \ref{fig:TVR-parallel-series} is the TVR. Note that the vertical
scale difference between the parallel and series connections is a
factor of 8 (18 dB) in the TVR and a factor of 64 the impedance, as
expected with the 8-ring segmented stack. The results show a small
improvement of the effective electromechanical coupling coefficient
in the series arrangement. Even more dramatic is the difference in
the higher frequency response. In the parallel arrangement, resonances
are observed in the electrical impedance near 40 kHz. However, in
the series experiments the next resonance is not observed until 68
kHz (see Figure \ref{fig:The-TVR-PMNPT}). This is similar to the
effect seen in Figure \ref{fig:Electrical-admittance-pmnpt}, although
that figure does not include the effects of water loading on the response.
A similar effect is seen in the TVR responses shown in Figure \ref{fig:TVR-parallel-series}.
Only minor differences are observed around the fundamental mode. However,
a 24 kHz difference is observed in the location of the null in the
TVR response. 

\noindent 

The FEA model does not agree perfectly with the position of the second
resonance in either the series or parallel configuration of the stack.
Reasons for that difference include an imperfect knowledge of the
piezoelectric parameters of the single crystal material, changes to
the material parameters that occurred during the depolarization and
repolarization to achieve the series connection, and the presence
of glue joints in the transducer. The agreement with measurement is
certainly good enough to understand that at least much of the performance
difference is explained by the series vs. parallel connection of the
piezoelectric rings.

\section{Conclusions}

This paper describes the effects of segmented stack design on the
performance of transducers built with high coupling single crystal
piezoelectric materials and compares those effects to a transducer
built with traditional PZT ceramics. One dimensional and finite element
models accurately predict the performance of the measured device.
The models show the change in the effective electromechanical coupling
coefficient of the complete transducer and the shifts in frequency
of higher order modes when the stacks are systematically segmented
into multiple pieces. Transducer designers using materials with high
electromechanical coupling coefficient to meet requirements for the
widest possible bandwidth should understand and consider these effects
when developing designs using high coupling piezoelectric materials.

\section*{Appendix}

\begin{figure}[b]
\noindent \begin{centering}
\includegraphics[width=2.5in]{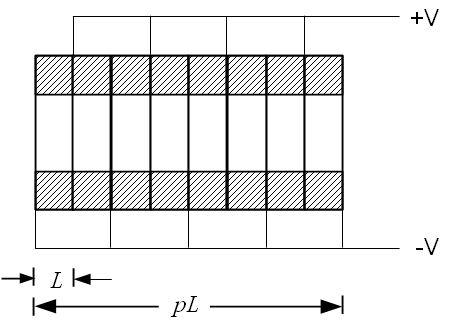}
\par\end{centering}

\noindent \caption{Stack dimensions as used in the Martin reference\cite{martin-segmented-33mode}.\label{fig:Stack-dimensions}}
\end{figure}

In references \cite{martin-segmented,martin-segmented-33mode} Martin
develops the conclusion that a ring stack of $p$ pieces each of thickness
$L$ can be modeled as a single element with length $pL$ with slightly
modified material parameters. (see Figure \ref{fig:Stack-dimensions})
In his derivations, Martin requires only that $L\ll\frac{\lambda}{2}$;
e.g. that the length of each ring segment is much less than a half
wavelength of vibrations in the piezoelectric material. This appendix
will show that Martin's conclusion actually requires that the length
constraint be $pL\ll\frac{\lambda}{2}$, and that further considerations
are necessary when the electromechanical coupling coefficient of the
piezoelectric material is higher than approximately 0.75. 

\noindent 

The constraint on the total length of the stack is generally understood
to be required for good performance by transducer designers. When
the stack length approaches a half wavelength, a mechanical resonance
within the stack will severely degrade the transducer performance.
Such long stacks would thus have been avoided in any case. Until recently,
the unstated constraint on the electromechanical coupling coefficient
was enforced by the fact that there were no available piezoelectric
materials with higher coupling. The current availability of single
crystal piezoelectric materials with electromechanical coupling coefficient
greater than 0.85 allow the error in the Martin derivation to be seen. 

\noindent 
\begin{figure}
\noindent \begin{centering}
\includegraphics[width=3in]{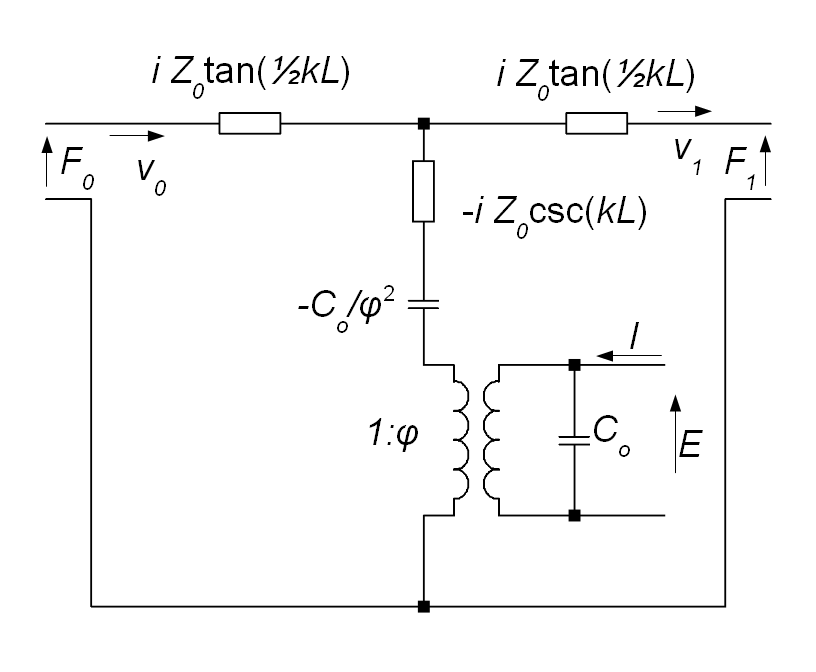}
\par\end{centering}

\caption{Analog circuit model for a single piezoelectric ring.\label{fig:Single-ring}}
\end{figure}

The Martin derivation correctly arrives at the 3-port model for a
single piezoelectric ring as the analog circuit of Figure \ref{fig:Single-ring}
where 

\begin{eqnarray}
C_{0} & = & \frac{\epsilon_{33}^{S}A}{L}=\frac{e_{33}^{T}A\left(1-k_{33}^{2}\right)}{L}\\
\varphi & = & \frac{A}{L}\left(\frac{\epsilon_{33}^{T}}{s_{33}^{E}}\right)^{\nicefrac{1}{2}}k_{33}\\
Z_{0} & = & \rho c_{L}A\\
k & = & \frac{\omega}{c_{L}}\\
c_{L} & = & \left(\rho s_{33}^{D}\right)^{\nicefrac{1}{2}}
\end{eqnarray}

\noindent and the equation for the impedance matrix described by this
circuit is

\noindent 
\[
\left[\begin{array}{c}
F_{0}\\
F_{1}\\
E
\end{array}\right]=\boldsymbol{Z}\left[\begin{array}{c}
v_{0}\\
v_{1}\\
I
\end{array}\right]
\]

\noindent with

\noindent 
\begin{equation}
\boldsymbol{Z}=\left[\begin{array}{ccc}
iZ_{0}\cot\left(\nicefrac{\text{kL}}{2}\right) & -iZ_{0}\csc\left(\text{\ensuremath{\nicefrac{\text{kL}}{2}}}\right) & \nicefrac{-i\varphi}{\omega C_{0}}\\
\\
-iZ_{0}\csc\left(\text{\ensuremath{\nicefrac{\text{kL}}{2}}}\right) & iZ_{0}\cot\left(\text{\ensuremath{\nicefrac{\text{kL}}{2}}}\right) & \nicefrac{-i\varphi}{\omega C_{0}}\\
\\
\nicefrac{-i\varphi}{\omega C_{0}} & \nicefrac{-i\varphi}{\omega C_{0}} & \nicefrac{-i}{\omega C_{0}}
\end{array}\right]\label{eq:exact-matrix-eq}
\end{equation}

\noindent where $F_{i}$ and $v_{i}$ are the mechanical force and
velocity at the two ends of the piece, and $E$ and $I$ are the voltage
and current at the electrical terminals. Note that this is the same
analog model used by Redwood, where Redwood recognizes that the components
with transcendental impedance functions can be represented by a mechanical
transmission line as in Figure~\ref{fig:Redwood-ckt}.

\begin{figure}
\noindent \begin{centering}
\includegraphics[width=3in]{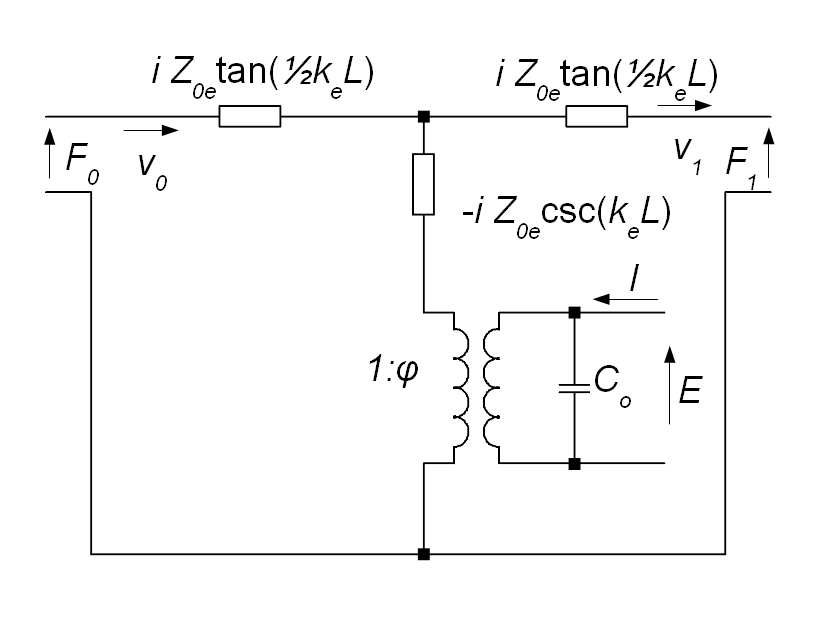}
\par\end{centering}

\caption{A simplified approximate analog circuit when $L\ll\nicefrac{\lambda}{2}$.\label{fig:single-ring-approx}}
\end{figure}

Next Martin chooses to consolidate the two components in the shunt
branch of the mechanical domain to give the approximation shown in
Figure~\ref{fig:single-ring-approx} where
\begin{eqnarray}
Z_{0e} & = & \rho c_{Le}A\label{eq:Z_oe}\\
k_{e} & = & \nicefrac{\omega}{c_{Le}}\label{eq:k_E}\\
c_{Le} & =c_{L} & \left(1-k_{33}^{2}\right)^{\nicefrac{1}{2}}\label{eq:c_Le}
\end{eqnarray}

\noindent Note that this is a low frequency approximation in which
the wave speed has been reduced to account for the negative compliance
present in the more accurate model of Figure~\ref{fig:Single-ring}.
This is an approximation that holds only when $L\ll\nicefrac{\lambda}{2}$.
The impedance matrix for this approximation becomes
\begin{equation}
\boldsymbol{Z}=\left[\begin{array}{ccc}
iZ_{0e}\cot\left(\nicefrac{k_{e}L}{2}\right) & -iZ_{0e}\csc\left(\nicefrac{k_{e}L}{2}\right) & \nicefrac{-i\varphi}{\omega C_{0}}\\
\\
-iZ_{0e}\csc\left(\nicefrac{k_{e}L}{2}\right) & iZ_{0e}\cot\left(\nicefrac{k_{e}L}{2}\right) & \nicefrac{-i\varphi}{\omega C_{0}}\\
\\
\nicefrac{-i\varphi}{\omega C_{0}} & \nicefrac{-i\varphi}{\omega C_{0}} & \nicefrac{-i}{\omega C_{0}}
\end{array}\right]\label{eq:approx_matrix-eq}
\end{equation}

\noindent 

Those familiar with transmission line modeling of 33-mode and 31-mode
piezoelectric pieces will notice that Martin's model for the ring
stack in Figure \ref{fig:single-ring-approx} has the form usually
ascribed to the 31-mode piece. One usually sees the negative compliance
in 33-mode pieces, and the lack of the negative compliance is often
taken as an implication that the 31-mode is being used. Martin's definition
is unusual in that the negative compliance has been combined into
the mechanical domain, whose material properties now depend on the
electromechanical coupling coefficient through the definitions in
Equations \ref{eq:Z_oe} through \ref{eq:c_Le}. This simplification
by Martin is correct, although most other references have followed
the lead of Redwood and kept the negative compliance as a separate
analog circuit component. 

In the final part of his derivation, Martin connects $p$ copies of
the analog circuit of Figure~\ref{fig:single-ring-approx} with their
mechanical ports cascaded end-to-end, and the electrical ports connected
in parallel. Martin's derived result for the ring stack is 
\begin{equation}
\boldsymbol{Z}=\left[\begin{array}{ccc}
iZ_{0e}\cot\left(\nicefrac{pk_{e}L}{2}\right) & -iZ_{0e}\csc\left(\nicefrac{pk_{e}L}{2}\right) & \nicefrac{-i\varphi}{\omega pC_{0}}\\
\\
-iZ_{0e}\csc\left(\nicefrac{pk_{e}L}{2}\right) & iZ_{0e}\cot\left(\nicefrac{pk_{e}L}{2}\right) & \nicefrac{-i\varphi}{\omega pC_{0}}\\
\\
\nicefrac{-i\varphi}{\omega pC_{0}} & \nicefrac{-i\varphi}{\omega pC_{0}} & \nicefrac{-i}{\omega pC_{0}}
\end{array}\right]\label{eq:approx_matrix-ring-stack}
\end{equation}

\begin{figure}
\begin{centering}
\includegraphics[width=3in]{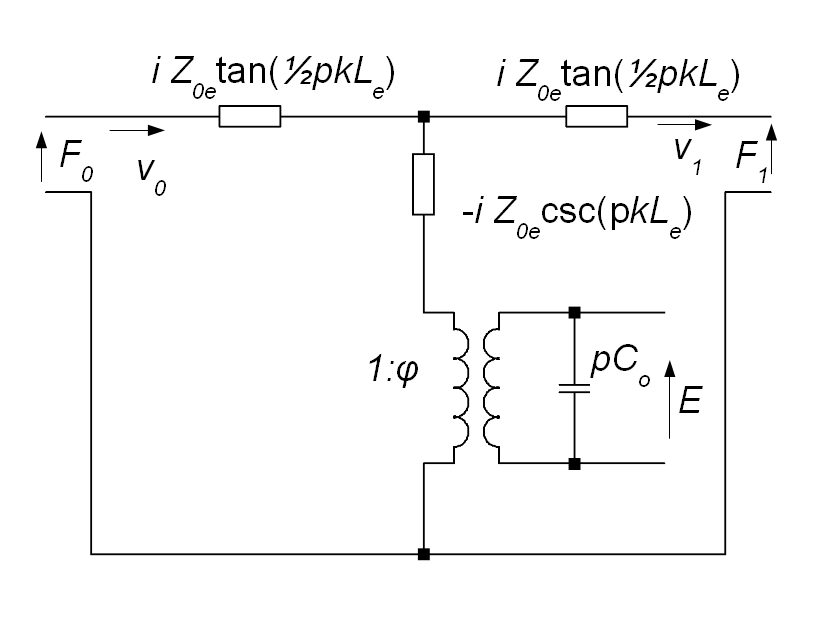}
\par\end{centering}

\caption{Martin's approximation for the piezoelectric ring stack.\label{fig:ckt-ring-stack}}
\end{figure}

\noindent From the similarity of Equations \ref{eq:approx_matrix-eq}
and \ref{eq:approx_matrix-ring-stack}, Martin has shown that the
analog circuit for the ring stack is that shown in Figure~\ref{fig:ckt-ring-stack}.
Martin claims that the only assumption in this derivation for the
ring stack is that the rings are identical and that the length of
each ring is small compared to the wavelength. The complete analysis
below will show that a broader set of restrictions is necessary for
stacks with a large number of rings and for stacks made from material
with high coupling coefficient.

\noindent 

Starting with Equation \ref{eq:approx_matrix-eq} with $k_{e}L_{E}\ll\pi$,
one may successively add a single additional ring $p$ times. At each
stage of this process, when the $n^{th}$ ring is added, the additional
fractional error terms $\delta_{n,1}$ and $\delta_{n,2}$ are added
as

\noindent {\small{
\begin{equation}
\boldsymbol{Z}=\left[\begin{array}{ccc}
iZ_{0e}\left(1-\delta_{n,1}\right)\cot\left(\nicefrac{pk_{e}L}{2}\right) & -iZ_{0e}\left(1+\delta_{n,2}\right)\csc\left(\nicefrac{pk_{e}L}{2}\right) & \nicefrac{-i\varphi}{\omega pC_{0}}\\
\\
-iZ_{0e}\left(1+\delta_{n,2}\right)\csc\left(\nicefrac{pk_{e}L}{2}\right) & iZ_{0e}\left(1-\delta_{n,1}\right)\cot\left(\nicefrac{pk_{e}L}{2}\right) & \nicefrac{-i\varphi}{\omega pC_{0}}\\
\\
\nicefrac{-i\varphi}{\omega pC_{0}} & \nicefrac{-i\varphi}{\omega pC_{0}} & \nicefrac{-i}{\omega pC_{0}}
\end{array}\right]\label{eq:matrix-with-error}
\end{equation}
}}{\small \par}

\noindent where

\noindent {\small{
\begin{eqnarray}
\delta_{n,1} & = & \frac{k_{33}^{2}}{nk_{e}L_{E}}\frac{\left(\sin nk_{e}L_{E}-n\sin k_{e}L_{E}\right)^{2}}{\left(n-1\right)\sin nk_{e}L_{E}-nk_{33}^{2}\mathrm{sinc}k_{e}L_{E}\sin\left[\left(n-1\right)k_{e}L_{E}\right]}\nonumber \\
 & \approx & \frac{k_{33}^{2}}{1-k_{33}^{2}}\frac{\left(n^{2}-1\right)\left(n+1\right)k_{e}^{4}L_{E}^{4}}{36}\mathrm{\,\,\,\,\,\,\, where\,\,}nk_{e}L_{E}\ll\pi\label{eq:delta-n-1-approx}\\
\delta_{n,2} & = & \frac{k_{33}^{2}}{nk_{e}L_{E}}\frac{\left(\sin nk_{e}L_{E}-n\sin k_{e}L_{E}\right)\left\{ \left(n-1\right)\sin nk_{e}L_{E}-n\sin\left[(n-1)k_{e}L_{E}\right]\right\} }{\left(n-1\right)\sin nk_{e}L_{E}-nk_{33}^{2}\mathrm{sinc}k_{e}L_{E}\sin\left[\left(n-1\right)k_{e}L_{E}\right]}\nonumber \\
 & \approx & \frac{k_{33}^{2}}{1-k_{33}^{2}}\frac{\left(n^{2}-1\right)\left(2n-1\right)k_{e}^{4}L_{E}^{4}}{36}\mathrm{\,\,\,\,\,\,\, where\,\,}nk_{e}L_{E}\ll\pi\label{eq:delta-n-2-approx}
\end{eqnarray}
}}{\small \par}

\noindent Note that for $1<n\leq p$ and $nk_{e}L_{E}<\nicefrac{\pi}{2}$,
all of these error terms are positive, and that they are monotonically
increasing with increasing $n$. To make this approximation, it is
necessary to stipulate that the total length of the ring stack is
short compared to the wavelength in the stack. It is not sufficient
that a single ring be short. Note also the fact that the electromechanical
coupling coefficient has a large impact on the error magnitude through
the term approximately proportional to $\nicefrac{n^{3}k_{33}^{2}}{\left(1-k_{33}^{2}\right)}$.
When PZT or other conventional piezoelectric material is used, and
if the effects of glue joints in the stack are considered, the effective
coupling coefficient is generally less than 0.6. This was the case
when the Martin papers were written. However, recent single crystal
piezoelectric materials may have coupling coefficients over 0.9, and
completed transducers using these materials can approach 0.85. This
difference in coupling coefficient can increase the magnitude of the
error by approximately an order of magnitude.

With modern computer hardware and software capabilities, there is
little need to use the Martin approximation for current designs. However,
designers should be aware of the limitations of Martin's approximation
to avoid its use in older analysis codes and to avoid introducing
it into new analyses. Designers using high coupling piezoelectric
materials must also guard against the previously safe intuitive understanding
that stack segmentation has no effect on the mechanical behavior of
a transducer employing a segmented stack. To the contrary, changes
to the usual conventions of stack segmentation may provide interesting
and useful modifications to the performance of transducers using high
coupling piezoelectric materials.

\pagebreak{}

\section*{Figure Captions}
\begin{enumerate}
\item The tonpilz transducer element used in this study has a stack of eight
PMN-PT crystal rings. (color online)
\item Analog circuit model for a single piezoelectric piece, after Redwood.
The mechanical domain includes the mechanical transmission line with
density $\rho$, sound speed $v$, cross sectional area $A$ and length
$L$. 
\item The plane wave model of the transducer is an interconnection of mechanical
transmission lines. The electrodes of the piezoelectric pieces are
connected electrically in parallel.
\item The piezoelectric stack is driven in four configurations, as eight
segments, four segments, two segments and one segment. Electrodes
remain in place for the eight rings, but not all are connected, as
shown.
\item The electrical admittance for transducers built with PZT-4 piezoelectric
material do not vary significantly in the frequency region of the
primary resonance and antiresonance .
\item Electrical admittance of the four segmented stacks using PMN-PT show
greater differences than with PZT in the frequency region of the primary
antiresonance, and especially at the higher mode resonance.
\item TVRs for the four transducer implementations using PZT-4.
\item The TVR for transducers built with PMN-PT single crystal piezoelectric
material show that segmentation may cause significant variation in
operating bandwidth. 
\item In water electrical impedance magnitude for a single crystal PMN-PT
based tonpilz element that was both measured and modeled in a) parallel
and b) series stack arrangement.
\item The measured Transmitting Voltage Response in water for a single tonpilz
element of single crystal PMN-PT. The element was measured and modeled
in a) series and b) parallel stack arrangement.
\item Stack dimensions as used in the Martin reference\cite{martin-segmented-33mode}.
\item Analog circuit model for a single piezoelectric ring.
\item A simplified approximate analog circuit when $L\ll\nicefrac{\lambda}{2}$.
\item Martin's approximation for the piezoelectric ring stack.\end{enumerate}

\end{document}